\documentclass[aps,prstab,preprint,groupedaddress]{revtex4-1}

\usepackage{amssymb}
\usepackage{amstext}
\usepackage{graphicx}

\begin{document}

\title{A Model for One-Dimensional Coherent Synchrotron Radiation including Short-Range Effects}

\author{Robert D. Ryne$^1$, Bruce Carlsten$^2$, Ji Qiang$^1$, Nikolai Yampolsky$^2$}
\affiliation{$^1$Lawrence Berkeley National Laboratory\\$^2$Los Alamos National Laboratory}

\date{\today}

\begin{abstract}
A new model is presented for simulating coherent synchrotron radiation (CSR) in one dimension.
The method is based on convolving an integrated Green function (IGF) with the longitudinal charge density.
Since it is based on an IGF, the accuracy of this approach is determined by how well one resolves the charge density and not by resolving the single particle wake function.
Since short-range wakefield effects are included analytically, the approach can be much more efficient than ordinary (non-IGF) approaches
in situations where the wake function and charge density have disparate spatial scales.
Two cases are presented: one derived from the full wake including short-range effects, and one derived from the asymptotic wake.
In the latter case the algorithm contains the same physics as others based on the asymptotic approximation,  but
requires only the line charge density and not its derivative.
Examples are presented that illustrate the limitations of the asymptotic-wake approximation, and that illustrate how microbunching
can increase the CSR fields by orders of magnitude depending on the microbunching wavelength.

\end{abstract}


\maketitle

\section{Introduction}
\begin{figure}[b]
\includegraphics[height=3.5in]{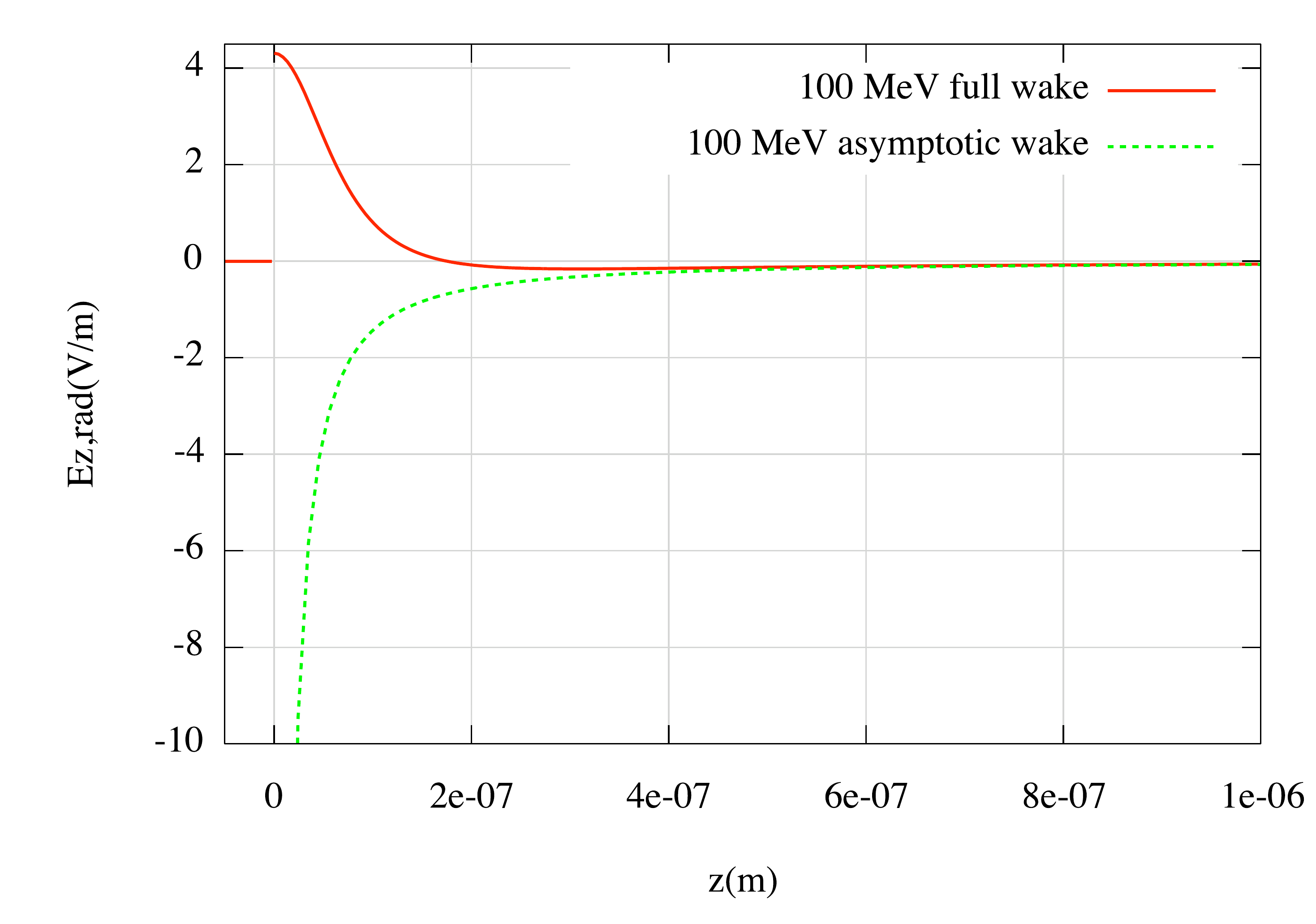}
\vspace{-0.3in}
\caption{The single-particle wake, $w(z)$, for a 100 MeV electron moving on a circle of radius $1~\rm m$,
assuming large $\gamma$, small angles, and steady state.
The full wake and its asymptotic approximation, $w\propto z^{-4/3}$, are shown.
The height of the full wake scales as $\gamma^4$ and the width scales as $1/\gamma^3$.
                }\label{fig0label}
\end{figure}

\vspace{-.2in}
Coherent synchrotron radiation (CSR) is one of the most important and difficult to model phenomena affecting future light sources and other lepton accelerators involving high intensity,
high energy, short bunches. It is also an important phenomena in certain astrophysical settings. Analyses of the radiation from a charged particle moving on a circle of radius $\rho$ were
performed by many authors including Schott \cite{schott} and later by Goldreich and Keely \cite{goldreich} and by Murphy, Krinsky, and Gluckstern \cite{murphy}.
As shown in \cite{murphy}, at relativistic velocities, in steady state, in the small angle approximation, and assuming free space boundary conditions,
the single particle wake function corresponding to the radiation component of the azimuthal electric field is, in MKS units,
$w=\left[{-1 \over 4\pi\epsilon_0}\right]{4 \over 3} {e \gamma^4 \over  \rho^2} \hat w(\mu),$
where the normalized wake function is,
\begin{equation}
\hat w(\mu)={d \hat\nu(\mu) \over d \mu},
\label{wisdmudnu}
\end{equation}
and where
\begin{equation}
\hat\nu(\mu)={9 \over 16} \left\{ {-2 \over \mu} + {1 \over \mu\sqrt{\mu^2+1}} \left( \Omega^{1/3}+ \Omega^{-1/3}   \right)  + {2 \over \sqrt{\mu^2+1}} \left( \Omega^{2/3}- \Omega^{-2/3}   \right)  \right\}.
\label{nudefine}
\end{equation}
In the above, $\Omega=\mu+\sqrt{\mu^2+1}$, $\mu={3 \gamma^3 \over 2 \rho} s$, and $s$ is the arc length between the observation point and the source charge.
Note that this is the wake for $\mu>0$; in the large $\gamma$ approximation the wake is zero for $\mu<0$.
Note that for small argument, $\hat w$ has the form,
$\hat w=1-{14\over 9}\mu^2+\cdots$, 
and, for large argument,
$\hat w\sim -{3/4 \over 2^{1/3}} \mu^{-4/3}+{9\over 8} \mu^{-2} + \cdots$~.
The single particle wake, $w$, is plotted as a function of $s$ in Fig.~\ref{fig0label} along with its asymptotic approximation.
Given the asymptotic form of $\hat w$ and the fact that $\mu\propto \gamma^3$ it is clear from these formulas (which are based on the large-$\gamma$ approximation)
that analyses involving just the asymptotic wake produce CSR fields that are independent of $\gamma$. In contrast, the short-range wake does produce energy-dependent effects.

The simulation of one-dimensional CSR effects usually involves computing the convolution of a wake function, $w$, with a numerically computed
line
charge density, $\lambda$, or its derivative $d\lambda/ds$.
If the single particle wake function depends only on the separation between the source and observation point, then the radiation electric field can be expressed as the convolution,
\begin{equation}
E(s)=\int{dz'}\lambda(s') w(s-s'),
\label{smoothconvolution}
\end{equation}
or
\begin{equation}
E(s)=-\int{ds'}{d\lambda(s')\over ds'} \nu(s-s'),
\label{smoothconvolutiondlambdadz}
\end{equation}
where $w(s)=d\nu/ds$ and where the integration is over a region for which $\lambda$ is zero at the boundaries.

A widely used method involves replacing $w$ with its asymptotic approximation \cite{saldin}. This results in a convolution of the form,
\begin{equation}
- {3/4\over 2^{1/3}}\int^s{ {ds' \over (s-s')^{4/3}}}\lambda(s'),
\label{fourthirdspower}
\end{equation}
or,
\begin{equation}
- {9/4\over 2^{1/3}}\int^s{ {ds' \over (s-s')^{1/3}}{d\lambda(s')\over ds'}}.
\label{onethirdpower}
\end{equation}
For this approach to be accurate, the characteristic length scale of the bunch must be much greater than $\rho/\gamma^3$. This is satisfied for typical parameters of existing accelerators,
but it is violated for highly compressed bunches, as well as for microbunched beams for proposed seeding schemes.
An exact model for 1D CSR is given in \cite{mayesandhoff} and \cite{saganetal} that involves a more complicated wake function but that, in contrast with Eq.~(\ref{nudefine}), does not assume small angles or small $\gamma$.
In general, the computation of CSR convolutions is challenging for two reasons. First, the behavior of the wake function near $s=0$ becomes increasingly localized and large with increasing $\gamma$.
Second, there can be a significant amount of numerical noise in the computed charge density, so methods that involve its derivative have to deal with this and usually involve careful smoothing.
Below a new approach is presented that does not involve differentiating the charge density, and which furthermore includes short-range wakefield effects.

\section{The integrated Green function approach to 1D CSR}
A simple discretization of Eq.~(\ref{smoothconvolution}) on a
grid with cell size $h$ leads to,
\begin{equation}
E_{k}=h \sum_{k'=1}^{k'_{max}}  \lambda_{k'}w_{k-k'},
\label{simplediscreteconvolution}
\end{equation}
where $\lambda_{k}$ and $w_{k-k'}$ denote the values of the line charge density and the wake function, respectively, defined on a grid of values $z_k$.
This simple discretization has a potentially significant pitfall:
It makes use of the wake function only at the grid points (even though it is known everywhere for certain problems where it can be determined analytically).
This can be problematic when $\lambda$ and $w$ have a disparate spatial variation, since, to achieve acceptable accuracy on a uniform grid,
a sufficiently fine grid is needed to resolve the smallest features of {\it both} $\lambda$ and $w$.
The use of a grid with variable grid spacing, or multiple grids of different resolution, can improve the efficiency of the algorithm, but with some additional algorithmic complexity.
In the case of the CSR, the wake is extremely narrow and strongly peaked near $z=0$.
Depending on the bunch length and its internal structure ({e.g.,} if it contains microbunches), the asymptotic wake, the short-range wake, or both, might be important.

There are numerous techniques for accurately and efficiently discretizing and evaluating Eq.~(\ref{smoothconvolution}).
Qiang described a method for approximating convolutions to arbitrary accuracy using the Newton-Cotes formula \cite{qiangnewtoncotes}.
For the CSR wake, Borland implemented a method of evaluating Eq.~(\ref{onethirdpower}) that is used in the \texttt{elegant} code and other codes \cite{borlandprstab2001}.
Carlsten implemented a method that includes some short-range radiation effects but requires solving a transcendental equation for the retarded times \cite{carlsten}.
The approach below does not require solving a transcendental equation, and furthermore contains all short-range radiation effects contained in the
single particle wake of Eq.~(\ref{wisdmudnu}).

Integrated Green functions (IGF's) provide a means to accurately and efficiently compute convolutions when certain integrals involving
the wake function can be obtained analytically or numerically \cite{oxford,prstabquasistaticmodel,abelletal}.
In the following we assume that the bunch length is so short compared to the bend radius that the longitudinal coordinate, $z$, can be used in place of the arc length, $s$.
For our purposes we use linear basis functions to approximate the charge density within a cell of the computational domain.
In that case Eq.~(\ref{smoothconvolution}) is approximated by,
\begin{equation}
E(z_k)={1\over h}\sum_{k'}  \int_{0}^{h}dz'~\left[(h-z')\lambda_{k'} + \lambda_{k'+1} z'\right] w(z_{k}-z_{k'}-z'),
\end{equation}
where the sum is over all values of $k'$ in the discretization of $\lambda$.
Shifting the indices in the last sum above and collecting terms,
the effective Green function is the coefficient of $h \lambda_{k'}$~,
\begin{equation}
E(z_k) =h \sum_{k'}  \lambda_{k'} w^{eff}_{k-k'},
\label{igf1dconvolution}
\end{equation}
where
\begin{equation}
w^{eff}_{\zeta}={1\over h^2}  \int_{0}^{h}dz'~\left[(h-z')w(\zeta-z')+ z' w(\zeta+h-z')\right] ,
\end{equation}
where $\zeta=z_{k}-z_{k'}$, and where we have used the notation $w^{eff}_{\zeta}$ to denote $w^{eff}_{k-k'}$.
Rearranging terms we obtain,
\begin{equation}
w^{eff}_{\zeta}={1\over h^2} \left[(h-\zeta) \int_{\zeta-h}^{\zeta}dz'~w(z') +  (h+\zeta) \int_{\zeta}^{\zeta+h}dz'~w(z')   +
 \int_{\zeta-h}^{\zeta}dz'~z' w(z') -  \int_{\zeta}^{\zeta+h}dz'~z' w(z')  \right] .
\label{igfbeforeparts}
\end{equation}
Integrating the last two terms by parts, and collecting terms, we obtain,
\begin{equation}
w^{eff}_{\zeta}={1\over h^2} \left[\int_{\zeta}^{\zeta+h}dz'~\nu(z') -  \int_{\zeta-h}^{\zeta}dz'~\nu(z')  \right] ,
\end{equation}
where $d\nu/dz=w$. Finally, let
\begin{equation}
\chi(z)=\int^z{dz'}\nu(z').
\label{chidefine}
\end{equation}
It follows that
\begin{equation}
w^{eff}_{\zeta}={1\over h^2} \Bigg(~\chi~\Big|_{\zeta+h} -2 \chi~\Big|_{\zeta}  +\chi~\Big|_{\zeta-h}   ~\Bigg) .
\label{copyformula121}
\end{equation}
Note, however, that because integration by parts was used to arrive at this result, it will not necessarily hold for all values of $\zeta$. In particular,
for many problems $w$ or $dw/dz$ is discontinuous at $z=0.$ In such cases $w^{eff}_0$ should be calculated separately.

The integral in Eq.~(\ref{chidefine}) can be done analytically for the 1D CSR wake of Eqs.~(\ref{wisdmudnu})-(\ref{nudefine}) (keeping in mind that the wake is zero for $\mu<0$.).
It follows that the effective Green function in Eq.~(\ref{igf1dconvolution}) is given by,
\begin{equation}
w^{eff}_{k}={1\over h^2} \Bigg\{~  \Bigg( \chi~\Big|_{k+1} -2 \chi~\Big|_{k}  +\chi~\Big|_{k-1} \Bigg) +\chi~\Big|_{0}\delta_{k,0}  ~\Bigg\} ~~~~~(k \ge 0),
\label{formula121}
\end{equation}
with $w^{eff}_k=0$ for $k<0$. In the above, $\chi_k$ is given for $k\ge 0$ by,
$\chi_k=\left[{-1 \over 4\pi\epsilon_0}\right] {16\over 27}{e \over \gamma^2}\hat\chi_k,$
where
\begin{equation}
\hat\chi_k={9 \over 16} \Bigg\{ 3  (-2\mu \Omega_k^{1/3}+\Omega_k^{2/3}+\Omega_k^{4/3}) 
+ \log{ [(1-\Omega_k^{2/3})/\mu]^2 \over (1+\Omega_k^{2/3}+\Omega_k^{4/3})        }     \Bigg\} ~~~~~(k \ge 0),
\label{chiformula}
\end{equation}
where $\Omega_k=\mu_k+\sqrt{\mu_k^2+1}$ and $\mu_k={3\gamma^3\over 2\rho}z_k$. Also, ${\hat\chi}_k=0$ for $k<0$.
Note that $\hat\chi$ is not singular, but, to evaluate it numerically for small argument one has the expansion,
$\hat\chi={9 \over 16} \{ 6-\log({27\over 4}) + {8 \mu^2 \over 9} - {56 \mu^4 \over 243} \cdots  \}$ for $\mu<<1.$

The preceding specifies the effective Green function obtained from the full wake.
If, instead of the full wake, its asymptotic expansion were used (as in Eq.~(\ref{fourthirdspower})),
the effective Green function is still given by Eq.~(\ref{formula121}) but with $\hat\chi_k$ replaced by $\hat\chi_k^{asymp}$, where
\begin{equation}
\hat\chi_k^{asymp}={27/8 \over 2^{1/3}} \mu_k^{2/3}~~~~~(k \ge 0),
\label{asympchiformula}
\end{equation}
with ${\hat\chi}_k^{asymp}=0$ for $k<0$.

\section{Examples}
\begin{figure}[h]
\begin{tabular}{ccc}
\includegraphics[height=1.5in]{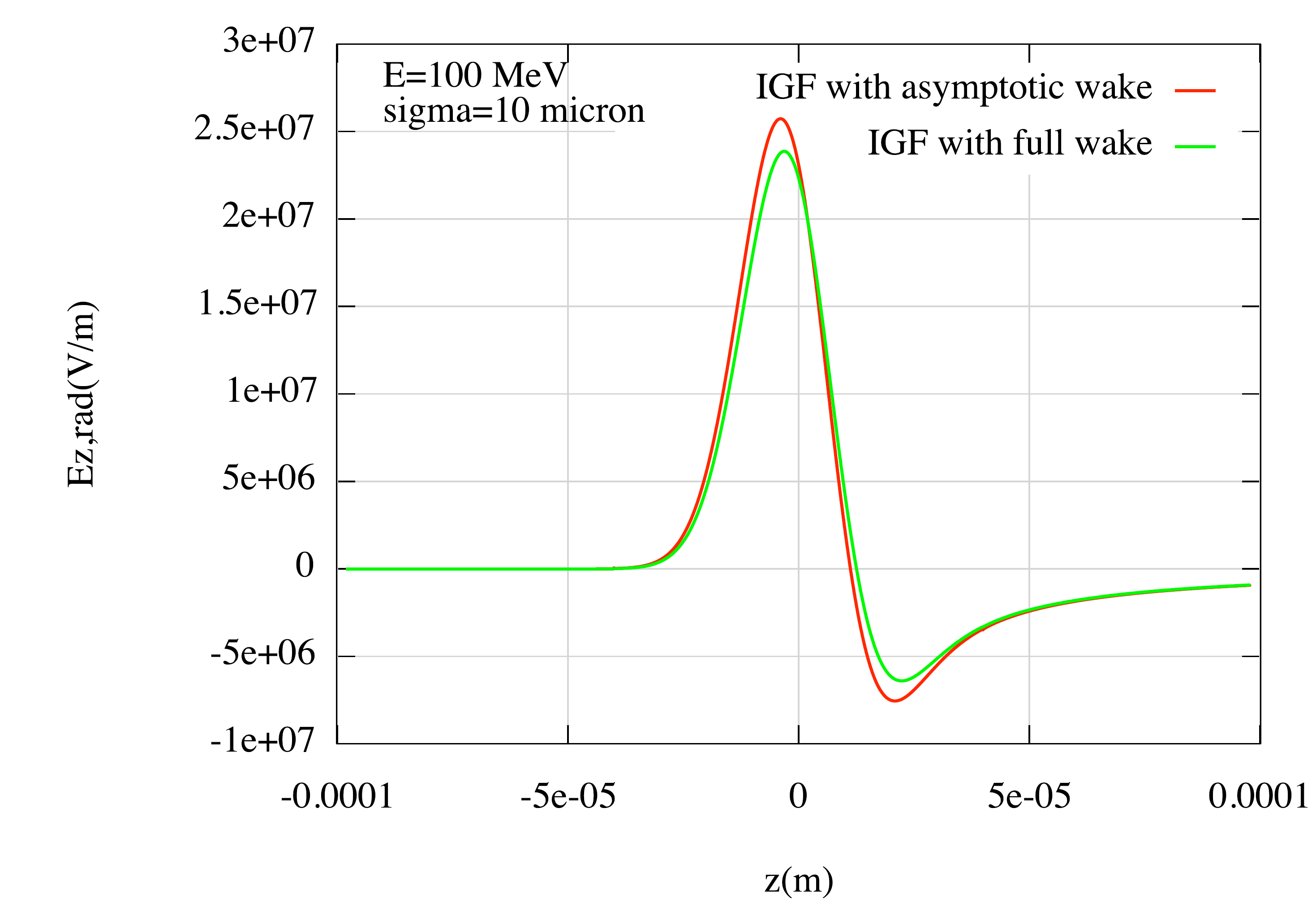} &
\includegraphics[height=1.5in]{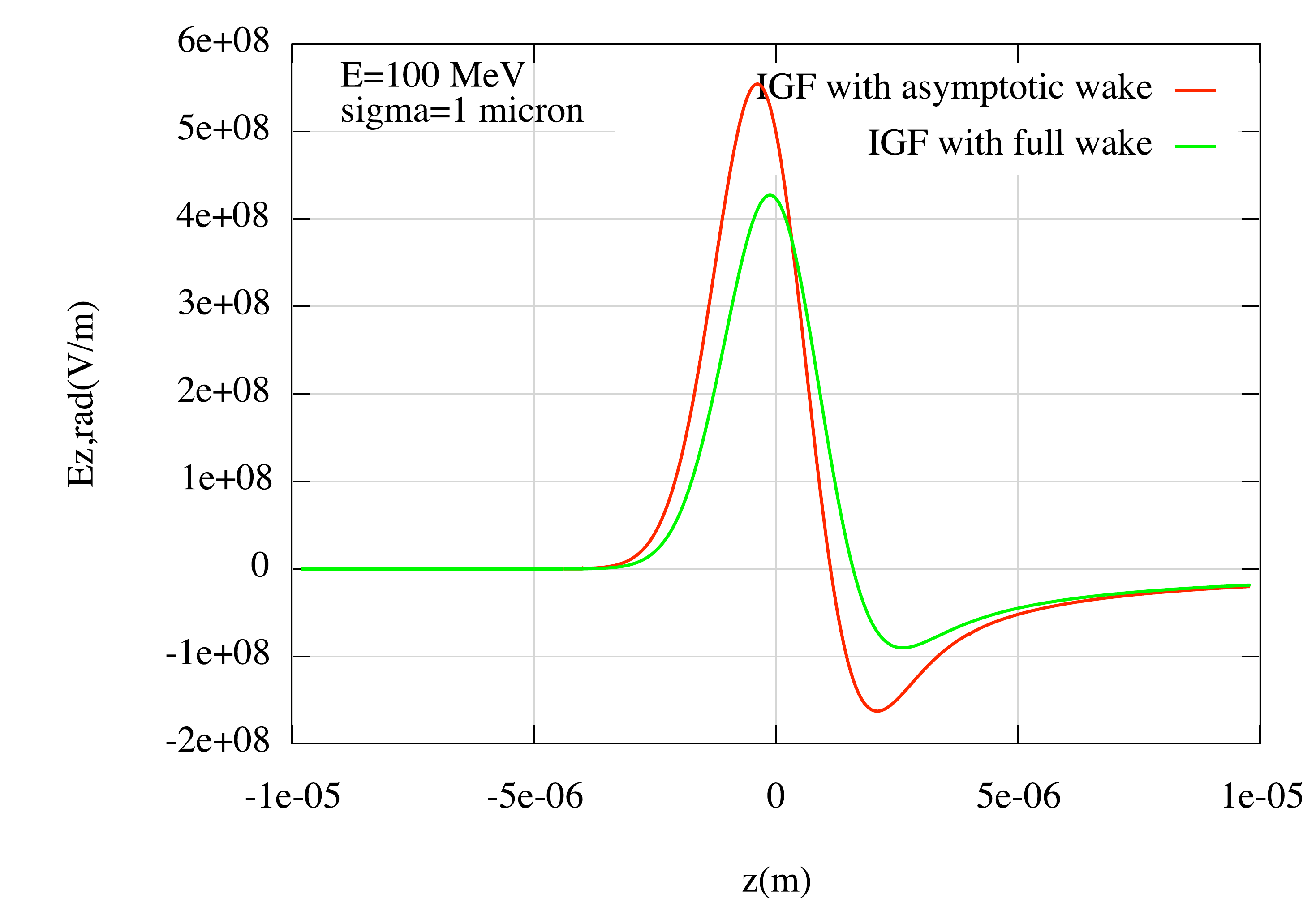} &
\includegraphics[height=1.5in]{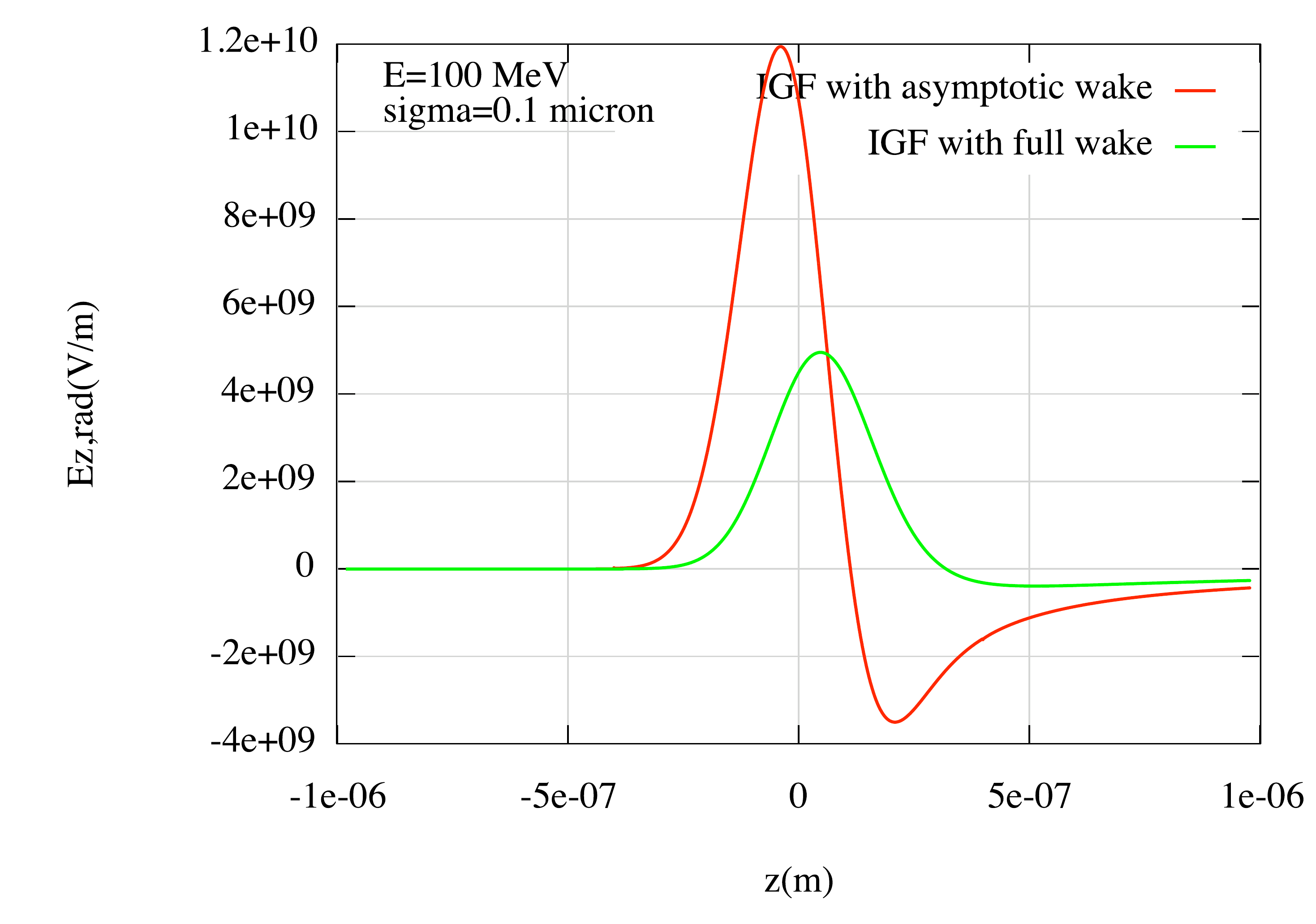} \\
\end{tabular}
\vspace{-0.3in}
\caption{Radiation component of the azimuthal electric field based on effective Green functions derived from the asymptotic wake and from the full wake. The beam is a 1 nC, 100 MeV Gaussian bunch with rms width $\sigma$.
Left: $\sigma=10~\rm micron$. Center: $\sigma=1~\rm micron$. Right: $\sigma=0.1~\rm micron$.
At a characteristic length scale of order $\sigma$, $\mu_{char}\sim 3\gamma^3 \sigma/(2\rho)=$ 114., 11.4, and 1.14 when $\sigma=10$, $1$, and $0.1~\rm micron$, respectively.
The condition $\mu_{char}>>1$ is violated in the right-hand figure, leading to significant error in the CSR calculation based on the asymptotic model (red curve).
                }\label{fig1label}
\end{figure}

\begin{figure}[h]
\includegraphics[height=2.25in]{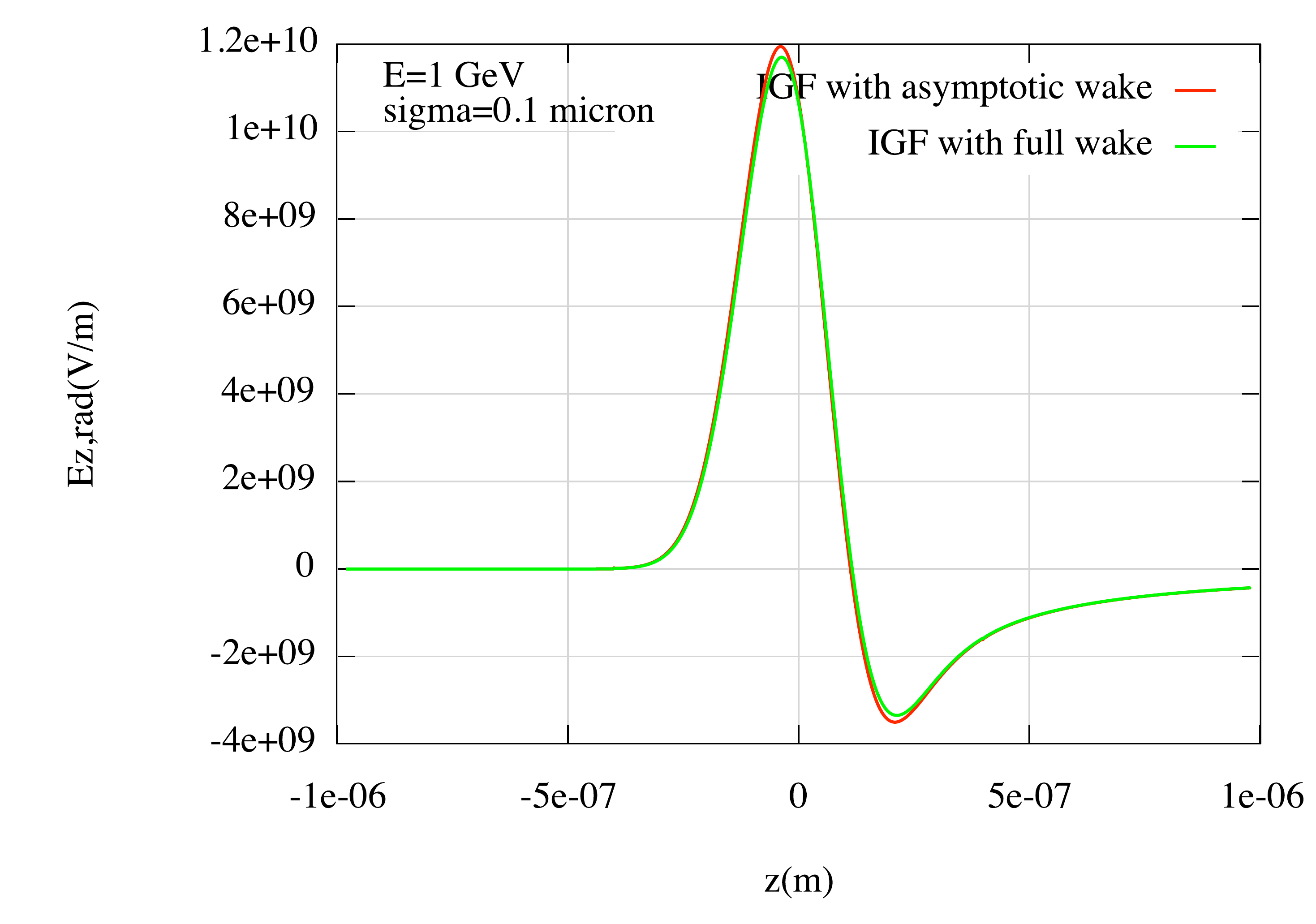}
\vspace{-0.3in}
\caption{Radiation component of the azimuthal electric field based on effective Green functions derived from the asymptotic wake and from the full wake.
The beam is a 1 nC, 1 GeV Gaussian bunch with rms width $\sigma=0.1~\rm micron$.
                }\label{fig2label}
\end{figure}

First consider a Gaussian line charge density with rms width $\sigma$.
The results that follow are proportional to the bunch charge, which we take to be 1 nC.
The bend radius of curvature is $\rho=1~\rm m$.
Figs.~\ref{fig1label} and \ref{fig2label} show the calculated radiation electric field using the IGF approach obtained from the asymptotic wake %(as in Eq.~(\ref{fourthirdspower}))
and from the full wake.
Fig.~\ref{fig1label} has  $E=100~\rm MeV$; the left, center, and right figures have $\sigma=10$, $1$, and $0.1~\rm micron$, respectively.
As is seen on the left, for the 10 micron case the asymptotic model gives results that are not significantly different from the exact model.
The differences are larger for the 1 micron case in the center. In Fig.~\ref{fig1label}, right, the result based on the asymptotic wake has a large error.
To understand this behavior, note that the characteristic length scale for this example is of order $\sigma$,
so the value of $\mu$ at the characteristic length scale is $\mu_{char}\sim 3\gamma^3/(2\rho)\sigma=$ 114., 11.4, and 1.14 when $\sigma=10$, $1$, and $0.1~\rm micron$, respectively.
In the 0.1 micron case the condition $\mu_{char}>>1$ is badly violated, resulting in significant error.
In contrast to Fig.~\ref{fig1label} right,  Fig.~\ref{fig2label} also shows results with $\sigma=0.1~\rm micron$ but at $E=1~\rm GeV$.
In this case, due to the higher energy $\mu_{char}\sim 1100$, so the asymptotic approximation is valid even for this short bunch length.

\begin{figure}[h]
\begin{tabular}{cc}
\includegraphics[height=2.25in,trim=15mm 5mm 0mm 5mm, clip]{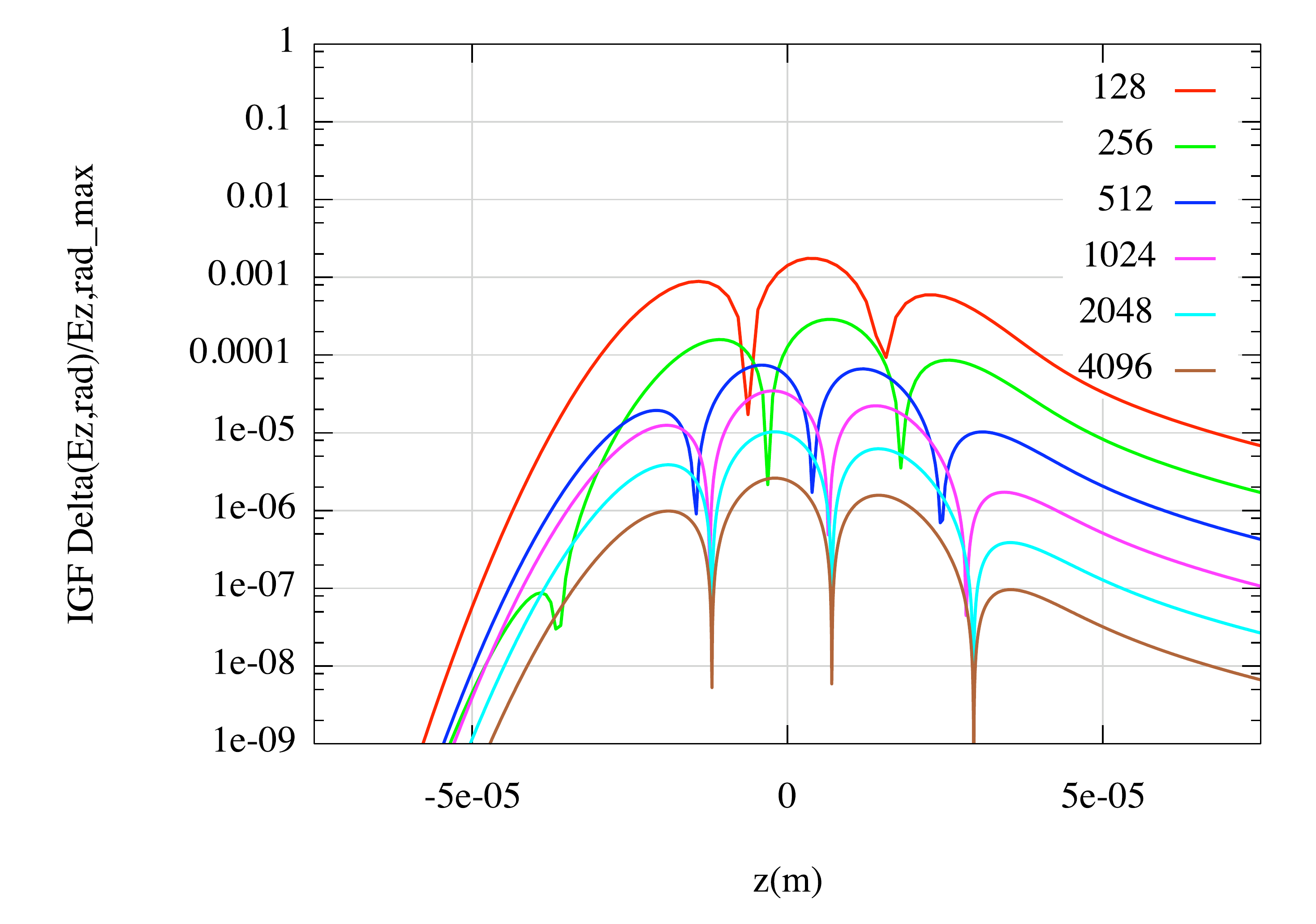} &
\includegraphics[height=2.25in,trim=5mm 5mm 0mm 5mm, clip]{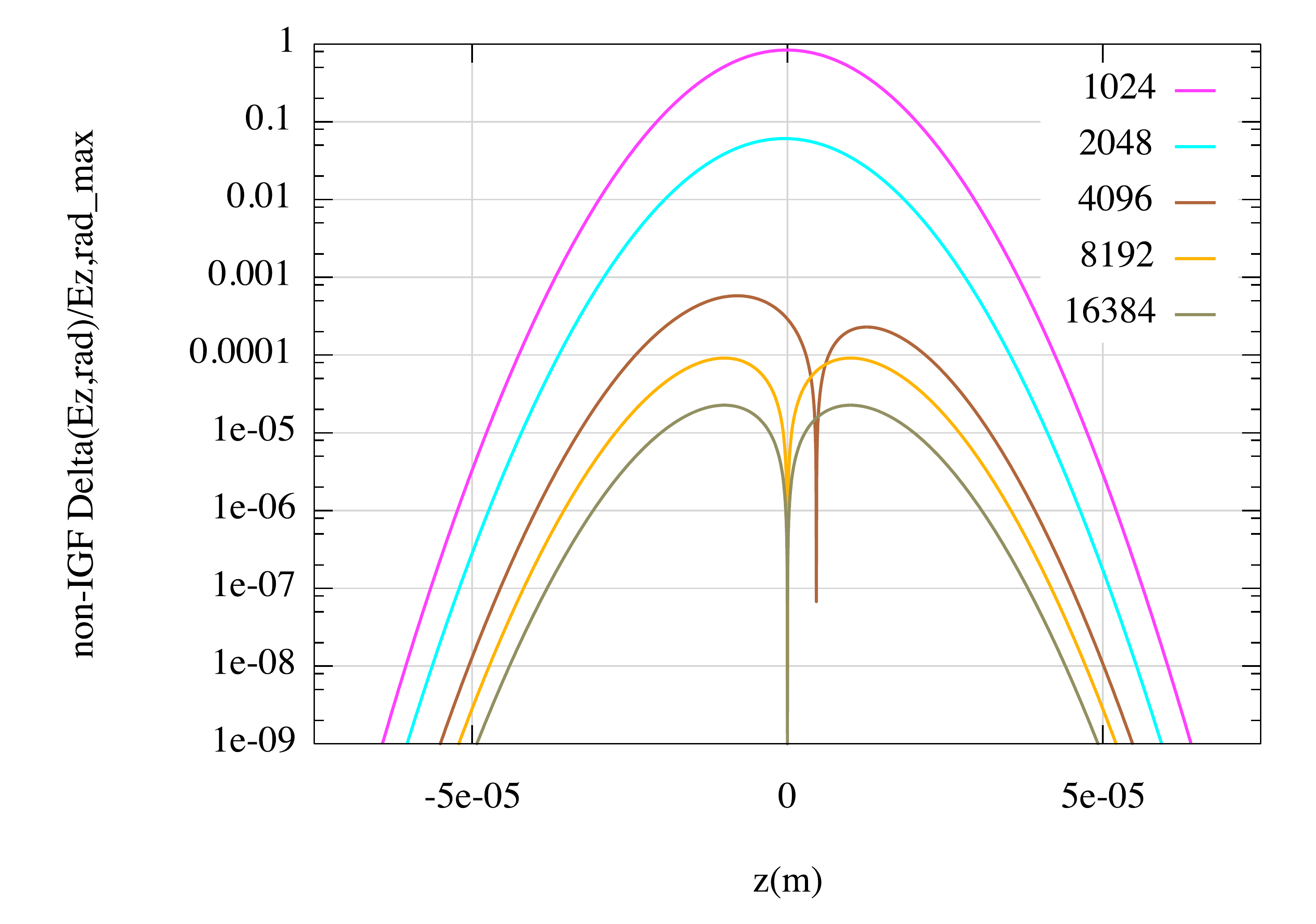} \\
\end{tabular}
\vspace{-0.3in}
\caption{Convergence test of the IGF method (left) and the non-IGF method (right) showing the relative error in the radiation component of the electric field for varying number of grid points.
The beam is a 1 nC, 100 MeV Gaussian bunch with rms width $\sigma=10~\rm micron$.
For these parameters the IGF method requires many fewer grid points (just 128) compared with the non-IGF method to achieve acceptable accuracy.
Alternatively, for these parameters the IGF method is up to several orders of magnitude more accurate than the non-IGF method at a given number of grid points.
                }\label{fig3label}
\end{figure}

The preceding demonstrated the ability of the IGF approach with the full wake function to capture the physics of short-range radiation effects. In addition, the IGF approach is more computationally efficient than the non-IGF approach
when the density varies slowly compared with the wake function. Returning to the case with $\sigma=10~\rm micron$, Fig.~\ref{fig3label} shows plots of the relative error in the radiation electric field,
$(E_{z,rad,num}-E_{z,rad,exact})/E_{z,rad,exact,max}$, where $E_{z,rad,num}$ is the numerical result, $E_{z,rad,exact}$ is the ``exact'' result based on the IGF approach with 500,000 grid points, and
where the difference has been normalized by the maximum value of the ``exact'' field.
Fig.~\ref{fig3label}, left shows the IGF, full wake results using 128, 256, 512, 1024, 2048, and 4096 grid points on a domain covering $\pm 10\sigma$.
The relative error is less than 0.1\% with just 128 points. This good accuracy is explained by the fact that 128 grid points are sufficient to reasonably resolve the Gaussian charge density over this domain;
it does not matter that 128 grid points are not sufficient to resolve the short-range structure of the wake.
Fig.~\ref{fig3label}, right shows the non-IGF, full wake results using 1024, 2048, 4096, 8192, and 16384 grid points.
In this case 4096 grid points are required to achieve an accuracy of 0.1\%  or less.
Taking a different viewpoint, at 1024 grid points the accuracy of the IGF approach is $.0035\%$ or less,
while that of the non-IGF approach is $85\%$ or less.
%Taking a different viewpoint, at 2048 grid points the accuracy of the IGF approach is roughly $1\times 10^{-5}$ or less,
%while the accuracy of the non-IGF approach is roughly $6\%$ or less.
%Taking a different viewpoint, at 4096 grid points the accuracy of the IGF approach is roughly $2.5\times 10^{-6}$ or less,
%while the accuracy of the non-IGF approach is roughly $.0005$ or less. This corresponds to a factor of 200 in accuracy.

\begin{figure}[h]
\begin{tabular}{l l}
\includegraphics[height=2.25in,trim=15mm 5mm 5mm 5mm, clip]{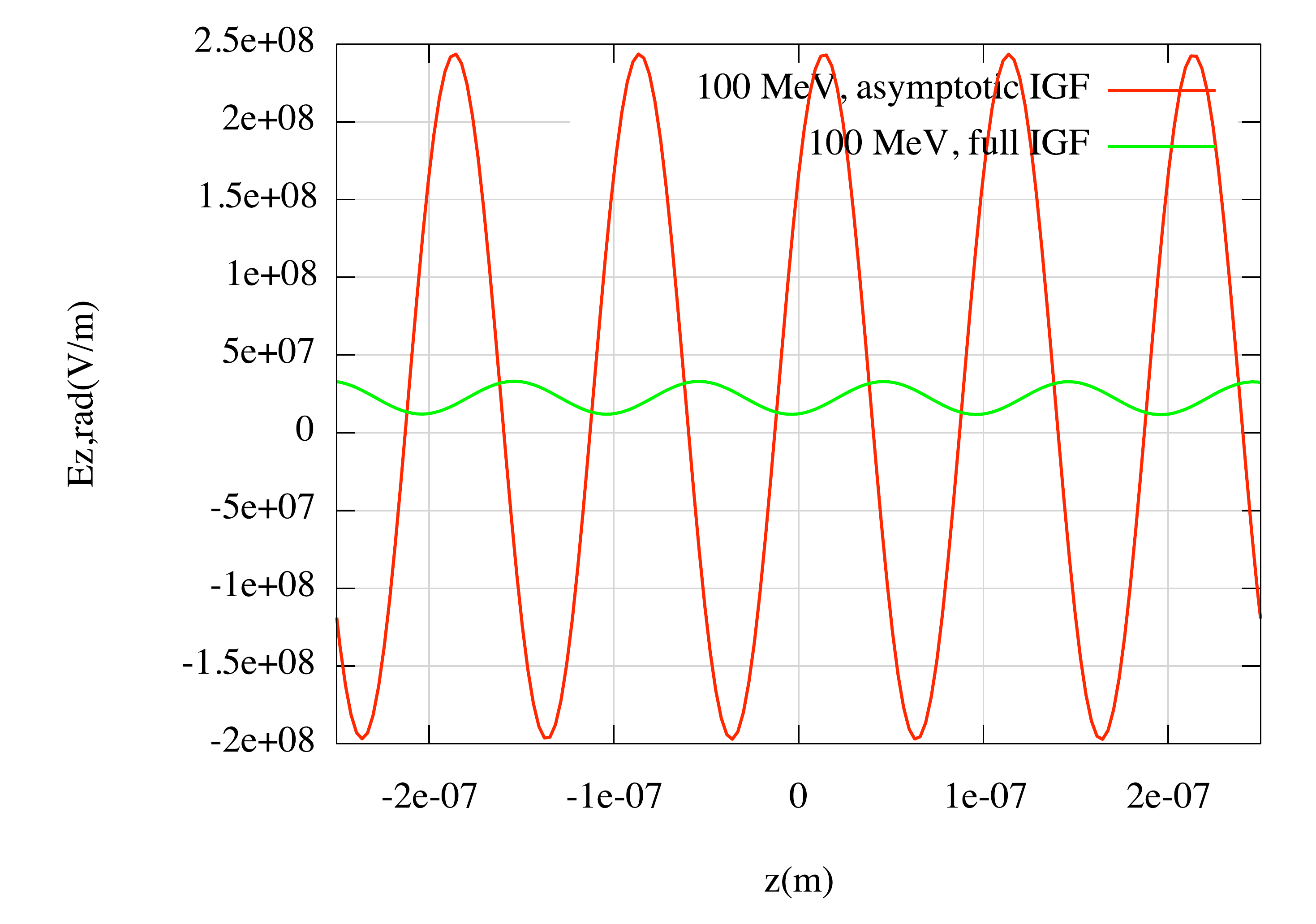} &
\includegraphics[height=2.25in,trim=5mm 5mm 5mm 5mm, clip]{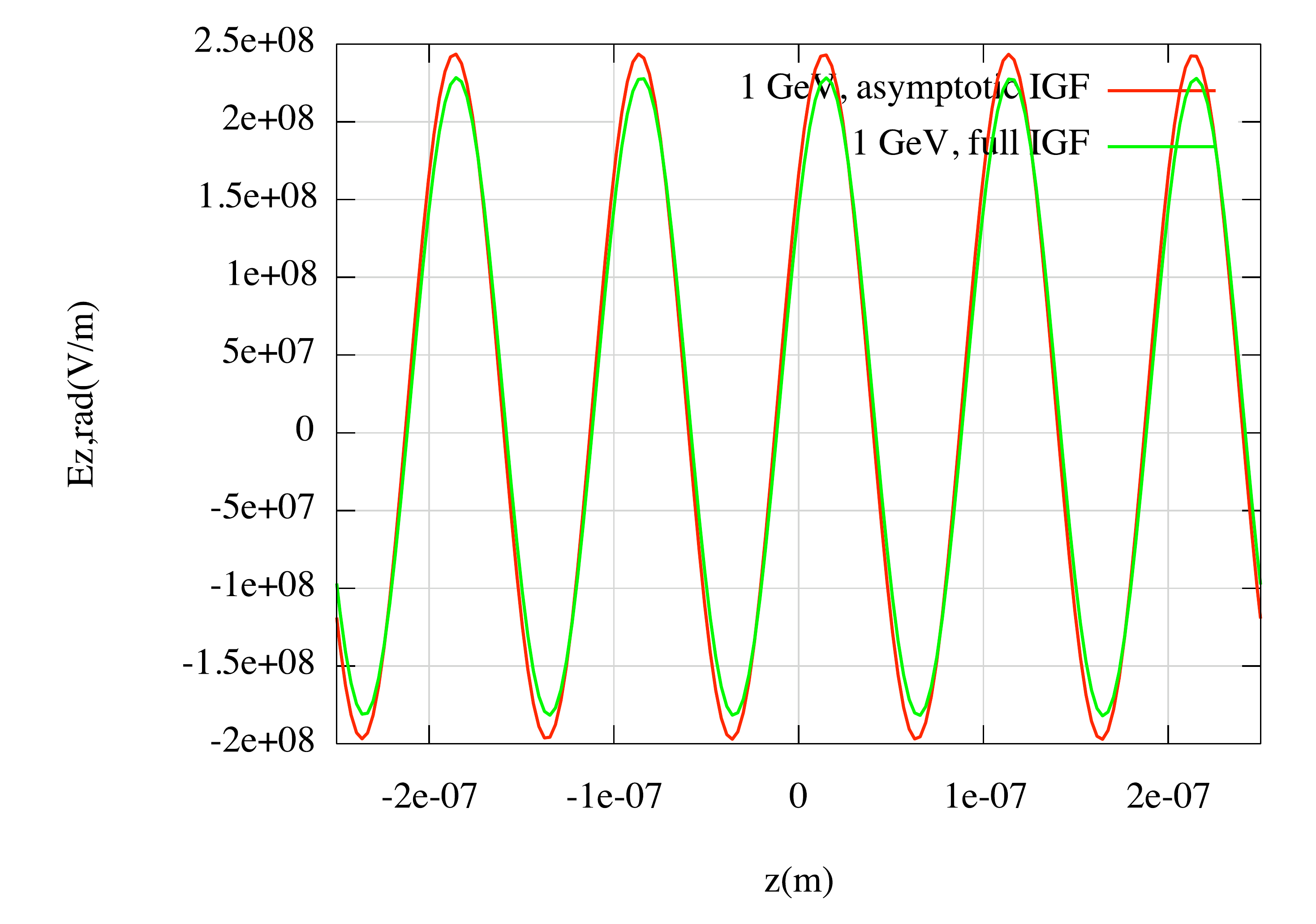} \\
\end{tabular}
\vspace{-0.3in}
\caption{Radiation component of the azimuthal electric field in a Gaussian bunch with sinusoidal microbunching.
The bunch rms width is $\sigma=10~\rm micron$, and the microbunching wavelength is $\lambda_{m}=100~\rm nm$. Left: 100 MeV. Right: 1 GeV.
The red curve on the left is wrong because, for these parameters at 100 MeV, the asymptotic wake is not appropriate since $\mu_{char}\sim 1$.
                }\label{fig4label}
\end{figure}

Lastly, consider a Gaussian line charge density with sinusoidal microbunching,
\begin{equation}
\lambda(z)={C\over 2}e^{-x^2/2\sigma^2}(
(1+f_{min}) + (1-f_{min})\sin(k_{m}z) )
),
\end{equation}
where $C$ is a normalization constant, $k_{m}=2\pi/\lambda_{m}$, $\lambda_{m}$ is the microbunching wavelength, and where $f_{min}$ controls the minimum depth of the modulation.
Fig.~\ref{fig4label} shows results with $\sigma=10~\rm micron$, $\lambda=100~\rm nm$, and $f_{min}=0.1$. 
The left figure is for an energy of 100 MeV, and the right figure is for 1 GeV. Both figures show IGF results with the full wake and with the asymptotic wake.
Particularly noteworthy is the difference in microbunching-driven field enhancement between 100 MeV and 1 GeV:
At 100 MeV the enhancement is approximately 35\%, at 1 GeV it is about 900\%.
Also, it is clear from the left plot that, for the 100 MeV parameters, the asymptotic IGF over-predicts the CSR field by a factor of 8, making the results unusable;
on the right at 1 GeV the asymptotic IGF gives results that are reasonably close to the full-wake results.
This is consistent with the earlier observation, namely, that the asymptotic-wake IGF can be used for the 1 GeV parameters (since $\mu>>1$), but
the full-wake IGF is required for the 100 MeV case.

\vspace{-0.3in}
\section{Discussion}
A new convolution-based method has been presented for computing the 1D CSR radiation electric field. Using an integrated Green function (IGF) approach,
an effective Green function was derived based on the 1D CSR wake. If the full single-particle wake of Eq.~(\ref{wisdmudnu}) is used,
the resulting effective Green function is given by Eqs.~(\ref{formula121}) and (\ref{chiformula}).
If the asymptotic wake is used, the resulting effective Green function is given by Eqs.~(\ref{formula121}) and (\ref{asympchiformula}).
In contrast with methods that use the derivative of the line charge density $d\lambda /dz$,
in this approach the discrete convolution is performed using $\lambda$.
If the analysis starts from the full CSR wake, the resulting convolution is able to capture both short- and long-range effects.
A numerical example demonstrated the ability of the full-wake IGF approach to model short-range effects when the asymptotic model is not appropriate.
The example also showed that the IGF approach has the potential to achieve accurate results, but with far fewer mesh points, $N$, than the non-IGF approach,
when $\lambda$ varies on a scale that is long compared with that of the single-particle wake.
The ability to use fewer mesh points has two consequences: first, an IGF code will run faster since the execution time scales as $2N\log(2N)$ for an FFT-based method,
and second, there will be less numerical noise since there will be more particles per cell.
(Note, however, that if the scale of the bunch microstructure less than or comparable to that of the short-range wake, then the mesh points needed to
resolve $\lambda$ will also resolve the short-range wake, and there might be no advantage to using an IGF.)
Lastly, an example with microbunching showed significant enhancement in a 1GeV bunch with microbunching at a wavelength of $100~\rm nm$.
This enhancement might (as in Fig.~\ref{fig4label}, right) or might not (as in Fig.~\ref{fig4label}, left) be modeled correctly in a code based on the asymptotic convolution, Eq.~(\ref{fourthirdspower}) or Eq.~(\ref{onethirdpower}), depending on the problem parameters.
The validity of the asymptotic model is most likely to be of concern at low energy, and in short bunches or in bunches with a microstructure.

The accuracy of the IGF approach depends on choosing a grid that is sufficient to resolve the charge density, not the single particle wake function.
Assuming that is the case, an IGF-based particle-in-cell code will usually produce acceptable results.
In contrast, a non-IGF code can produce unusable results, even when the grid is fine enough to resolve the charge density,
if the Green function is not well-resolved by the grid.
In general IGF codes tend to be more robust than their non-IGF counterparts since they are less sensitive to resolution issues associated with the grid spacing.

The approach described here does not require computing the derivative of the line charge density, $d\lambda/dz$.
This is desirable given the loss of accuracy in computing the numerical derivative of a possibly noisy, numerically derived function.
In principle one could make use of $d\lambda/dz$ in an IGF code by applying the IGF technique to Eq.~(\ref{smoothconvolutiondlambdadz}).
But there is no obvious advantage to doing this since, whether one uses Eq.~(\ref{smoothconvolution}) or (\ref{smoothconvolutiondlambdadz}), the singularity is handled analytically.
From a programming standpoint there is never any need to compute $d\lambda/dz$.
Instead one simply deposits charge on a 1D grid and performs the discrete convolution given by Eqs.~(\ref{formula121}) and (\ref{chiformula}).
In regard to computational effort, numerical convolutions can be computed rapidly
by turning the discrete convolution into a cyclic discrete convolution that can be handled using Fast Fourier Transforms methods (see the Appendix of \cite{rynearxiv}).

The examples presented illustrate the dependence of the radiation field on both energy and bunch length scale:
On one hand, the short-range wake grows as $\gamma^4$, so there is potential for huge field enhancement at high energy, with important implications for microbunched beams.
On the other hand, if $\mu\sim\gamma^3 s/\rho>>1$ then the short-range portion of the wake is irrelevant, the asymptotic portion dominates, and the radiation field is independent of $\gamma$.

We have presented an IGF-based method to calculate 1D CSR. In light of that it is reasonable to ask, under what conditions is the 1D approximation likely to be valid?
An approximate criterion can be obtained as follows:
The condition that the radiation cones of two electrons separated by a transverse distance $x$ overlap is, for small angles,
$x\sim \rho\theta\psi_{ret}$ where $\rho$ is the bend radius, $\theta$ is the opening angle, and where $\psi_{ret}$ is the retarded angular position.
For frequencies smaller than the critical frequency $\omega_c=(3/2) \gamma^3 c/\rho$, the coherent radiation is primarily within the angle $\theta=(1/\gamma)(\omega_c/\omega)^{1/3}$.
The dominant radiation is at frequencies of order of the inverse bunch length, $\omega\sim 2\pi c/\sigma$. 
The range of retarded angles is related to the bunch length by $\psi_{ret}\approx (24\sigma/ \rho)^{1/3}$.
It follows that the overlap condition for a bunch of rms transverse size $x$ and rms length $\sigma$ is $x\sim \rho (\sigma/\rho)^{2/3}$,
and the 1D approximation is likely to be valid as long as $x$ is well below this value.
We have found this criterion to be consistent with results from a particle-based, massively parallel Lienard-Wiechert code \cite{rynecarlstenyampolsky}.

Lastly, two phenomena should be mentioned that are not included in this analysis. First, we have discussed just the radiation component of the field, not the Coulomb component.
Since the Coulomb field varies as $1/r^2$, it can't be handled in the one-dimensional approximation since the integral in  Eq.~(\ref{smoothconvolution}) would be divergent.
It can, however, be handled using a two-dimensional model.
Second, we have not discussed shot noise effects. Simulations using a massively parallel Lienard-Wiechert code have shown that
shot noise effects can be a major consideration near 1 GeV and above, even in bunches with several billion electrons \cite{rynecarlstenyampolsky}.
In general, classical CSR effects in a bunch containing a finite number of particles involve a smooth component and a stochastic component.
The analysis presented here describes an efficient way to model the smooth component of the radiation field in 1D, and to include
both short- and long-range phenomena if needed.

In conclusion, the method described by Eqs.~(\ref{formula121}) and (\ref{chiformula}) provides a robust method for computing 1D CSR effects.
Since it includes both short- and long-range effects, it will produce more reliable results than the asymptotic model after the microbunching instability sets in,
({\it i.e.}, after small-scale structures develop),
resulting in simulations that provide more reliable estimates of the microbunching growth rate.
Given the simplicity of the algorithm, it's numerical efficiency, and the added benefit that it does not require numerically differentiating the line charge density,
this provides an attractive new method for modeling 1D steady-state CSR effects.
\vspace{-.2in}
\section*{Acknowledgements}
\vspace{-.2in}
This research was supported by the Laboratory Directed Research and Development program of Los Alamos National Laboratory,
and by the Office of Science of the US Department of Energy, Office of Basic Energy Sciences, Accelerator and Detector Research and Development program.
This research used resources of the National Energy Research Scientific Computing Center, which is supported by the Office of Science of the U.S. Department of Energy under Contract No. DE-AC02-05CH11231.

\bibliographystyle{model1-num-names}
\bibliography{<your-bib-database>}

\end{document}